\documentclass{article}

\usepackage{tikz, pgf, pgfplots}
\usetikzlibrary
	{
		arrows, automata, chains, datavisualization.formats.functions, fit, positioning, quantikz, shapes
	}

\usepackage{amsthm}
\usepackage{amssymb}
\usepackage[bb = boondox]{mathalfa}
\usepackage{dsfont}
\usepackage{mathtools}
\usepackage{multicol}
\usepackage{physics}
\numberwithin{equation}{section}

\usepackage{yquant, braket}
\usetikzlibrary{quantikz, fit}

\usepackage[a4paper, left = 2.5cm, right = 2.5 cm, top = 2.5cm, bottom = 3.0 cm]{geometry}

\usepackage{xcolor}
\usepackage{varwidth}
\usepackage[breakable, skins, theorems]{tcolorbox}
\usepackage{graphicx}
\usepackage{hyperref}
\hypersetup
{
	colorlinks,
	linkcolor = {WordRed!75!black},
	citecolor = {WordBlueDarker25},
	urlcolor = {WordAquaDarker50}
}
\usepackage{nicematrix}
\usepackage{colortbl}
\usepackage{float}
\usepackage{cellspace}
\usepackage{makecell}
\usepackage[boxed, ruled, vlined]{algorithm2e}
\usepackage{appendix}
\definecolor{MyLightRed}{RGB}{244, 213, 245}
\definecolor{WordRed}{RGB}{255, 0, 102}
\definecolor{RedDarkLightest}{HTML}{ff0088}
\definecolor{RedDarkLight}{HTML}{ea005f}
\definecolor{RedPurple}{HTML}{aa007f}
\definecolor{Purple}{HTML}{911146}
\definecolor{WordLightGreen}{RGB}{140, 214, 192}
\definecolor{WordGreen}{RGB}{0, 176, 80}
\definecolor{GreenLightest}{HTML}{00ffa0}
\definecolor{GreenLighter1}{HTML}{00b383}
\definecolor{GreenLighter2}{HTML}{00aa7f}
\definecolor{GreenDark}{HTML}{225522}
\definecolor{GreenTeal}{HTML}{008080}
\definecolor{WordIceBlue}{RGB}{223, 227, 229}
\definecolor{MyVeryLightBlue}{RGB}{211, 245, 247}
\definecolor{WordBlueVeryLight}{RGB}{0, 176, 240}
\definecolor{WordBlueLight}{RGB}{0, 112, 192}
\definecolor{WordBlueDark}{RGB}{46, 116, 181}
\definecolor{WordBlueDarker}{RGB}{31, 78, 121}
\definecolor{WordBlueDarker25}{RGB}{54, 96, 146}
\definecolor{WordBlueDarker50}{RGB}{36, 64, 98}
\definecolor{WordBlueDarkest}{RGB}{0, 32, 96}
\definecolor{WordBlue}{RGB}{19, 65, 99}
\definecolor{MyBlue}{RGB}{0, 64, 128}
\definecolor{MyDarkBlue}{RGB}{0, 51, 102}
\definecolor{BlueVeryDark}{HTML}{222255}
\definecolor{WordAquaLighter80}{RGB}{218, 238, 243}
\definecolor{WordAquaLighter60}{RGB}{183, 222, 232}
\definecolor{WordAquaLighter40}{RGB}{146, 205, 220}
\definecolor{WordAquaDarker25}{RGB}{49, 134, 155}
\definecolor{WordAquaDarker50}{RGB}{33, 89, 103}
\definecolor{WordVeryLightTeal}{RGB}{223, 236, 235}
\definecolor{WordLightTeal}{RGB}{160, 199, 197}
\definecolor{WordDarkTealLighter80}{RGB}{207, 223, 234}
\definecolor{WordDarkTeal}{RGB}{72, 123, 119}
\definecolor{WordDarkerTeal}{RGB}{48, 82, 80}
\definecolor{WordTurquoiseLighter80}{RGB}{209, 238, 249}

\title{QKD based on symmetric entangled Bernstein-Vazirani}

\author{
	Michael Ampatzis$^1$ and Theodore Andronikos$^1$\\
	$^1$Department of Informatics, Ionian University, \\
	Corfu, Greece; \{p16abat, andronikos\}@ionio.gr \\
}

\begin{document}

\maketitle

\begin{abstract}
	This paper introduces a novel entanglement-based QKD protocol, that makes use of a modified symmetric version of the Bernstein-Vazirani algorithm, in order to achieve a secure and efficient key distribution. Two variants of the protocol, one fully symmetric and one semi-symmetric, are presented. In both cases, the spatially separated Alice and Bob share multiple EPR pairs, one qubit of the pair each. The fully symmetric version allows both parties to input a secret key from their respective location and, finally, acquire in the end a totally new and original key, an idea which was inspired by the Diffie-Hellman key exchange protocol. In the semi-symmetric version, Alice sends her chosen secret key to Bob (or vice versa). Furthermore, their performance against an eavesdroppers attack is analyzed. Finally, in order to illustrate the operation of the protocols in practice, two small scale but detailed examples are given.

\end{abstract}

\section{Introduction} \label{sec:Introduction}

Upon the course of the last century, the scientific community experimented with different ideas and forms of computation, trying to harness the power of nature and create machines that allowed us to process immeasurable amounts of information in mere seconds, thus radically changing the world around us in the span of a few decades. However, in the present era classical computers are reaching a point where it will be infeasible to substantially enhance their efficiency due to the physical limitations of transistors. This started a new incentive to resurrect previous attempts concerning research of new types of computation. Out of all the different proposals for a viable substitute to classical computing, undoubtedly the most promising of them all is quantum computation, mainly due to the fact that it allows the exploitation of the most fundamental properties of physics.

\subsection{Related work}

As technology comes closer to the realization of this goal, it appears that certain profound adaptations regarding different branches of computer science need to take place in order to achieve a smoother transition from the classical to the quantum era. One of the most important such branches is the field of cryptography, due to the vulnerability of the current security algorithms against quantum computers \cite{Shor1999}, \cite{Grover1996}. This inherent weakness in the modern security protocols and the race for building a resilient security infrastructure against quantum computers \cite{chen2016report} before they become a reality, were the two catalysts that resulted in a schism of the field into two sub-fields, which are based on two different philosophies and ideologies. The first sub-field, known as post-quantum cryptography or quantum-resistant cryptography, relies on the complexity of mathematics as its security basis and is an attempt to develop cryptographic systems that are secure against both quantum and classical computers, and can also be interpreted within the already existing communications protocols and networks. The second sub-field, which is called quantum cryptography, is being built upon the implementation of the properties of quantum mechanics and, thus, takes advantage of nature's own fundamental laws in order to achieve security.

The sub-field of quantum cryptography, on which the primary interest of the current paper lies upon, has seen some enormous growth of both theoretical and practical nature. Two landmark papers, the BB84 protocol \cite{Bennett1984} and the E91 protocol \cite{Ekert1991}, were the first papers that proved that key distribution between two parties relying on the properties of quantum mechanics was possible. These two protocols have established the two schemes that all quantum key distribution (QKD) protocols are based on, the \emph{prepare-and-measure-based scheme} and the \emph{entanglement-based scheme}. After the publications of these two protocols, a plethora of interesting proposals for different QKD protocols based on these two schemes were suggested, further expanding the field on a theoretical level. While at the same time, some truly remarkable real life implementations of some protocols were demonstrated as in \cite{bennett1989experimental, elliott2005current, elliott2018darpa, peev2009secoqc, sasaki2011field, liao2017satellite}. These implementations have demonstrated that quantum cryptography is not just a mere theoretical experiment, but a possible reality in the near future.

Over the last few years, there was a noticeable increase in the effort to find new viable applications for well-known quantum algorithms, such as the Deutsch-Jozsa algorithm \cite{Deutsch1992}, the Bernstein-Vazirani algorithm \cite{Bernstein1997} and Simon’s periodicity algorithm \cite{Simon1997}. Many of these proposals have been made in the field of quantum cryptography, by utilizing these algorithms as viable QKD protocols \cite{Nagata2017, Nagata2017c, Nagata2018}. Motivated from these attempts, this paper proposes two novel variants of an entanglement-based QKD protocol that makes use of the Bernstein-Vazirani algorithm. The novelty of this work lies on the fact that it uniquely combines the following ingredients. Entanglement is an integral part of the protocols because the corresponding qubits in Alice and Bob's input registers are maximally entangled. Thus, the proposed protocols exhibit all the inherent advantages that an entanglement-based QKD protocol provides in terms of security against an eavesdropper, as first demonstrated in the E91 protocol \cite{Ekert1991}. Additionally, the Bernstein–Vazirani algorithm \cite{Bernstein1997}, a fast and useful quantum algorithm that guarantees the creation of the key using just one application of the appropriate function, is utilized in a critical manner. Furthermore, the fully symmetric variant is inspired by the Diffie-Hellman idea \cite{Diffie1976} of deriving the final key from a random combination of two separate keys. This idea is not just cosmetic, as the ability to obtain a key that neither Alice or Bob know from the start, adds an additional layer of security, further improving the strength of the protocol. Finally, the proposed protocol can be implemented in two versions: the fully symmetric version and the semi-symmetric one. In the fully symmetric variant, both Alice and Bob can input their tentative secret keys from their respective locations and acquire in the end a totally new and original key. In the semi-symmetric one, Alice (alternatively Bob) constructs the secret key that she (or he) communicates securely to the other party.

The protocol is described as a quantum game, which, despite the rather playful name, it is another noteworthy field that has emerged due to our transition to the quantum era and is used to address difficult and interesting problems within the quantum realm. This approach was chosen in an effort to make the presentation more mnemonic and easier to follow, due to the close connection that both fields share and the fact that any cryptographic situation can be conceived as a game between the two fictional heroes Alice and Bob, who play the roles of two remote parties that are trying to communicate, and the enemy Eve who tries to eavesdrop the conversation, a case which becomes apparent with the quantum game of coin tossing and the BB84 protocol \cite{Bennett1984}, \cite{Bennett2014} and references therein. This situation has been generalized in \cite{Aharon2010} to quantum dice rolling. For the reader striving for a more rounded understanding about the connection of the two fields, one can start with the two important works in the field of quantum game theory dating back to the year 1999, which were instrumental for the creation of the field: Meyer's PQ penny flip game \cite{Meyer1999}, which can be regarded as the quantum analogue of the classical penny flip game, and the introduction of the Eisert-Wilkens-Lewenstein scheme \cite{Eisert1999} that is widely used in the field. Regarding the PQ penny flip game, some recent results can be found in \cite{Andronikos2018} and \cite{Andronikos2021a}. As for the Eisert-Wilkens-Lewenstein scheme, it proved fruitful in providing many interesting results. For example, it led to quantum adaptations of the famous prisoners' dilemma in which the quantum strategies are better than any classical strategy (\cite{Eisert1999}), as well as extensions of the classical repeated prisoners' dilemma conditional strategies to a quantum setting (\cite{Giannakis2019}).

\subsection{Organization}

The paper is structured as follows. Section \ref{sec:Introduction} provides a brief introduction to the subject and gives the most relevant references. Section \ref{sec:Preliminaries} introduces and explains the tools used for the formulation of the protocols in this article. Section \ref{sec:QKD based on Symmetric Entangled B-V} presents and thoroughly analyzes the fSEBV and sSEBV protocols, so that their functionality can be completely understood. Section \ref{sec:Examples Illustrating the Operation of the Protocols} contains two detailed examples, one for each protocol, to demonstrate their operation. Finally, Section \ref{sec:Conclusions} provides a brief summary and discussion of the proposed protocols.

\section{Preliminaries} \label{sec:Preliminaries}

\subsection{Quantum entanglement and Bell states} \label{subsec:Quantum entanglement and Bell states}

Quantum entanglement is one of the fundamental principles of quantum mechanics and can be described mathematically as the linear combination of two or more product states. The Bell states are specific quantum states of two qubits, sometimes called an EPR pair, that represent the simplest examples of quantum entanglement. From the perspective of quantum computation, an EPR pair can be produced by a circuit with two qubits, in which a Hadamard gate is applied to the first qubit and subsequently a CNOT gate is applied to both qubits. These states can be elegantly described by the following equation taken from \cite{Nielsen2010}.

\begin{align} \label{eq:Bell States General Equation}
	\ket{\beta_{x,y}} = \frac{ \ket{0} \ket{y} + (-1)^x \ket{1} \ket{\Bar{y}} } {\sqrt{2}} \ ,
\end{align}

where $\ket{\Bar{y}}$ is the negation of $\ket{y}$.

In a more detailed manner, the Bell states can be described as follows.

\begin{minipage}[b]{0.45 \textwidth}
	\begin{align} \label{eq:Bell State Phi +}
		\ket{\Phi^{+}} = \ket{\beta_{00}} = \frac{ \ket{0} \ket{0} + \ket{1} \ket{1} } {\sqrt{2}}
	\end{align}
\end{minipage} 
\hfill
\begin{minipage}[b]{0.45 \textwidth}
	\begin{align} \label{eq:Bell State Phi -}
		\ket{\Phi^{-}} = \ket{\beta_{10}} = \frac{ \ket{0} \ket{0} - \ket{1} \ket{1} } {\sqrt{2}}
	\end{align}
\end{minipage} 

\begin{minipage}[b]{0.45 \textwidth}
	\begin{align} \label{eq:Bell State Psi +}
		\ket{\Psi^{+}} = \ket{\beta_{01}} = \frac{ \ket{0} \ket{1} - \ket{1} \ket{0} } {\sqrt{2}}
	\end{align}
\end{minipage} 
\hfill
\begin{minipage}[b]{0.45 \textwidth}
	\begin{align} \label{eq:Bell State Psi -}
		\ket{\Psi^{-}} = \ket{\beta_{11}} = \frac{ \ket{0} \ket{1} - \ket{1} \ket{0} } {\sqrt{2}}
	\end{align}
\end{minipage} 

The main advantage of quantum entanglement is that if one qubit of the pair gets measured, then the other will collapse immediately despite the distance between the two. This unique characteristic of quantum entanglement can be utilized on quantum key distribution as firstly described by Ekert in the E91 protocol. Therefore, in order to achieve quantum key distribution, multiple EPR pairs will be needed. For this reason, the mathematical representation of multiple EPR pairs will be expedient. If one starts with the entangled Bell state $\ket{\Phi^{+}}$, which can be cast as

\begin{align} \label{eq:Extended Bell State Phi +}
	\ket{\Phi^{+}}
	=
	\frac{1}{ \sqrt{2} }
	\left( \ket{0}_{A} \ket{0}_{B} + \ket{1}_{A} \ket{1}_{B} \right) \ ,
\end{align}

some easy computations show that

\begin{align} \label{eq:N Bell States Phi +}
	\ket{\Phi^{+}}^{\otimes n}
	&=
	\frac{1}{ \sqrt{2^n} }
	\sum_{\mathbf{x} \in \{ 0, 1 \}^n}
	\ket{\mathbf{x}}_{A} \ket{\mathbf{x}}_{B} \ ,
\end{align}

which will be required in the presentation of Section \ref{sec:QKD based on Symmetric Entangled B-V}.

\subsection{A brief description of the Bernstein-Vazirani algorithm} \label{subsec:A brief description of the Bernstein-Vazirani algorithm}

Regarded as one of the earliest quantum algorithms, along with the Deutsch-Josza algorithm and Simon's periodicity algorithm, the Bernstein-Vazirani algorithm, firstly introduced by Ethan Bernstein and Umesh Vazirani, can be considered as useful extension of the Deutsch-Josza algorithm, due to the fact that it was directly inspired from it and shared multiple common characteristics on both structure and implementation. Yet, despite the similarities, it has proved its value by demonstrating that the superiority of a quantum computer can be successfully utilized for more complex problems than the Deutsch-Josza problem.

The Bernstein-Vazirani problem can be described as the ensuing game between two players, namely Alice and Bob, who are spatially separated. Alice in Athens is corresponding with Bob in Corfu using letters. Alice starts the game by selecting a number $x$ from $0$ to $2^n-1$ and mails its binary $n$-bit representation $\mathbf{x}$ to Bob. After Bob receives this message, he calculates the value of some function

\begin{align} \label{eq:The Bernstein Vazirani I}
	f: \{ 0, 1, \dots, 2^{n} - 1 \} \rightarrow \{ 0, 1 \} \ ,
\end{align}

and replies with the result, which is either $0$ or $1$. The rules of the game dictate that Bob must use a function $f_{ \mathbf{s} } ( \mathbf{x} )$, where $\mathbf{s} = s_{n - 1} \dots s_{1} s_{0}$ and $\mathbf{x} = x_{n - 1} \dots x_{1} x_{0}$ are $n$-bit binary numbers representing integers in the range $0, 1, \dots, 2^{n} - 1$, such that

\begin{align} \label{eq:The Bernstein Vazirani II}
	f_{ \mathbf{s} } ( \mathbf{x} ) = \mathbf{s} \cdot \mathbf{x} \bmod 2 \ .
\end{align}

The inner product modulo 2 is defined as

\begin{align} \label{eq:Inner Product Modulo 2}
	\mathbf{s} \cdot \mathbf{x} \bmod 2 = s_{n - 1} x_{n - 1} \oplus \dots \oplus s_{0} x_{0} \ ,
\end{align}

where $\oplus$ is the exclusive-or operator. Therefore, the function is guaranteed to return the bitwise product of Alice's input $\mathbf{x}$ with a secret key $\mathbf{s}$ that Bob has chosen. Alice's goal in this game is to determine with certainty the secret key $\mathbf{s}$ that Bob has picked, corresponding with him as little as possible. How fast can she succeed?

In the \emph{classical} version of this problem, Alice can find the secret key $\mathbf{s}$ by taking advantage of the nature of the function $f_{ \mathbf{s} } ( \mathbf{x} )$ and, in particular, by sending Bob the inputs shown in Table \ref{tbl:Alice must communicate with Bob $n$ times}.

\begin{table}[H]
	\centering
	\renewcommand{\arraystretch}{1.5}
	\begin{tabular}{ >{\centering\arraybackslash} m{3.5 cm} !{\vrule width 1.25 pt} >{\centering\arraybackslash} m{3.5 cm} }
		\Xhline{4\arrayrulewidth}
		\multicolumn{2}{c}{The evolution of the Bernstein-Vazirani game}
		\\
		\Xhline{\arrayrulewidth}
		Alice's input $\mathbf{x}$
		&
		Bob's response
		\\
		\Xhline{3\arrayrulewidth}
		$\mathbf{x} = 1 0 \dots 0 0$
		&
		$s_{n - 1}$
		\\
		\Xhline{\arrayrulewidth}
		$\vdots$
		&
		$\vdots$
		\\
		\Xhline{\arrayrulewidth}
		$\mathbf{x} = 0 0 \dots 1 0$
		&
		$s_{1}$
		\\
		\Xhline{\arrayrulewidth}
		$\mathbf{x} = 0 0 \dots 0 1$
		&
		$s_{0}$
		\\
		\Xhline{4\arrayrulewidth}
	\end{tabular}
	\renewcommand{\arraystretch}{1.0}
	\caption{Alice must communicate with Bob $n$ times in order find the secret key $\mathbf{s}$.}
	\label{tbl:Alice must communicate with Bob $n$ times}
\end{table}

In that way, Alice will reveal a bit of the string $\mathbf{s}$ (the bit $s_{i}$) with each query she sends. For example, with $\mathbf{x} = 1 0 \dots 0$ she can obtain the most significant bit of $\mathbf{s}$, with $\mathbf{x} = 0 1 \dots 0$ she will find the next most significant bit of $\mathbf{s}$, and by continuing the same procedure all the way to $\mathbf{x} = 0 0 \dots 1$ she will finally manage to reveal the entire string $\mathbf{s}$. Despite, the efficiency of this method, Alice is still limited by sending only one query to Bob at a time. Therefore, the best possible classical scenario requires from her to correspond with Bob at least $n$ times, in order for her to succeed in her goal.

By observing the core attributes of the aforementioned game, we can divide it into the following three big steps, which are:

\begin{itemize}
	\item	Alice provides an input,
	\item	Bob applies the function $f_{ \mathbf{s} } ( \mathbf{x} )$, and
	\item	after multiple repetitions of the previous two steps, Alice is finally able to reveal the secret key $\mathbf{s}$.
\end{itemize}

From these steps, it is quite obvious that the game can easily be exploited, if the two players were able to implement certain tools from quantum mechanics, namely the use of quantum superposition, in order to allow Alice to send every possible input of $\mathbf{x}$ at once and the ability to manipulate these inputs with the use of a unitary transform. Thus, if Alice and Bob were able to exchange information with the use of qubits instead of classical bits, and Bob was using a unitary transformation $U_f$ instead of a function $f_{ \mathbf{s} } ( x )$, then Alice would be able to achieve her goal with only one communication.

The quantum version of the Bernstein-Vazirani algorithm, can be described by the following quantum game. The game, initially starts with Alice preparing two quantum registers, one of size $n$ to store her query in and one of size $1$, which she will send to Bob to store his answer in. We will refer to these registers as Alice's input and output registers, respectively. Next, she applies the Hadamard gate to every qubit, in order to acquire the even superposition state of each register and then she sends both registers to Bob. Right after Bob receives the contents of the registers, he applies the unitary transform $U_f$ and sends them back to Alice. In the final stage of the game, Alice concludes the algorithm by measuring her input register and obtaining the secret key $\mathbf{s}$. The whole process of the game, is summarized in Figure \ref{fig:The B-V Algorithm} below.

\begin{figure}[H]
	\centering
	\begin{tikzpicture} [scale = 1.25]
		\begin{yquant}
			nobit AUX1;
			qubits {$\ket{0}^{\otimes n}$} IR;	
			nobit AUX2;
			qubit {$\ket{1}$} OR;				
			nobit AUX3;
			[ name = S0 ]
			barrier (AUX1, AUX3);
			hspace {0.3 cm} IR;
			align IR, OR;
			box {$H^{\otimes n}$} IR;
			box {$H$} OR;
			[ name = S1 ]
			barrier (AUX1, AUX3);
			hspace {0.3 cm} IR;
			box {$ \quad U_{f_{A}} \quad $} (IR, AUX2, OR);
			[ name = S2 ]
			barrier (AUX1, AUX3);
			hspace {0.3 cm} IR;
			box {$H^{\otimes n}$} IR;
			[ name = S3 ]
			barrier (AUX1, AUX3);
			hspace {0.3 cm} IR;
			measure IR;
			hspace {0.3 cm} IR;
			output {$\ket{ \mathbf{s} }$} IR;
			\node [ below = 3.0 cm ] at (S0) {$\ket{\psi_{0}}$};
			\node [ below = 3.0 cm ] at (S1) {$\ket{\psi_{1}}$};
			\node [ below = 3.0 cm ] at (S2) {$\ket{\psi_{2}}$};
			\node [ below = 3.0 cm ] at (S3) {$\ket{\psi_{3}}$};
		\end{yquant}
		\node (BVAlgorithm) at (4.0, 0.75){\textbf{The Bernstein-Vazirani algorithm}};
	\end{tikzpicture}
	\caption{This figures gives a schematic representation of the Bernstein-Vazirani algorithm.}
	\label{fig:The B-V Algorithm}
\end{figure}
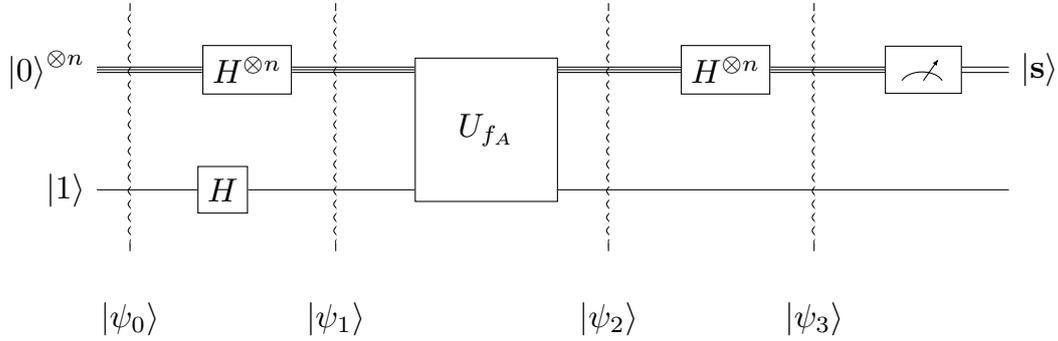

Now, in order to get a better understanding about the nature of the algorithm, let us examine the evolution of the quantum states more closely. First, Alice starts with the initial state

\begin{align} \label{eq:BV Initial State}
	\ket{\psi_0}
	=
	\ket{0}^{\otimes n} \ket{1} \ .
\end{align}

The $n$ qubits of her input register are all prepared at state $\ket{0}$ and the qubit of the output register is prepared at state $\ket{1}$. Next, Alice applies the Hadamard transform to both registers and the state becomes

\begin{align} \label{eq:BV Phase 1}
	\ket{\psi_1}
	=
	\frac{1}{\sqrt{2^n}}
	\sum_{\mathbf{x} \in \{ 0, 1 \}^n} \ket{\mathbf{x}}
	\left( \frac{\ket{0} - \ket{1}}{\sqrt{2}} \right) \ .
\end{align}

The derivation of the previous equation is based on the fact that

\begin{align} \label{eq:Walsh-Hadamard Transform on Basis Ket $0^n$}
	H^{\otimes n} \ket{0}^{\otimes n}
	=
	\frac{1}{\sqrt{2^n}}
	\sum_{\mathbf{x} \in \{ 0, 1 \}^n }^{}
	\ket{\mathbf{x}} \ ,
\end{align}

a standard result in the literature (for its derivation see \cite{Mermin2007} or \cite{Nielsen2010}). At this point the input register is in an even superposition of all possible states and the output register is in an evenly weighted superposition of $\ket{0}$ and $\ket{1}$. Thus, Alice is now ready to send both registers to Bob so he may apply the function $f_{ \mathbf{s} } ( x )$ by using

\begin{align} \label{eq:BV Oracle}
	U_{f} : \ket{\mathbf{x}, y} \rightarrow \ket{\mathbf{x}, y\oplus f(\mathbf{x})} \ ,
\end{align}

which results in the next state

\begin{align} \label{eq:BV Phase 2 - I}
	\ket{\psi_2}
	=
	\frac{1}{\sqrt{2^n}}
	\sum_{\mathbf{x} \in \{ 0, 1 \}^n}
	(-1)^{f(\mathbf{x})}
	\ket{\mathbf{x}}
	\left( \frac{\ket{0} - \ket{1}}{\sqrt{2}} \right) \ .
\end{align}

In view of (\ref{eq:The Bernstein Vazirani II}), (\ref{eq:BV Phase 2 - I}) becomes

\begin{align} \label{eq:BV Phase 2 - II}
	\ket{\psi_2}
	=
	\frac{1}{\sqrt{2^n}}
	\sum_{\mathbf{x} \in \{ 0, 1 \}^n}
	(-1)^{\mathbf{s} \cdot \mathbf{x}}
	\ket{\mathbf{x}}
	\left( \frac{\ket{0} - \ket{1}}{\sqrt{2}} \right) \ ,
\end{align}

which is the state returned back to Alice.

Let us now recall the following well-known equation that gives in a succinct form the result of the application of the Hadamard transformation to an arbitrary $n$-qubit basis ket $\ket{ \mathbf{x} }$ (see \cite{Mermin2007} or \cite{Nielsen2010}).

\begin{align} \label{eq:Walsh-Hadamard Transform on Arbitrary Basis Ket $x$}
	H^{\otimes n} \ket{ \mathbf{x} }
	=
	\frac{1}{\sqrt{2^n}}
	\sum_{\mathbf{z} \in \{ 0, 1 \}^n }^{}
	(-1)^{\mathbf{z} \cdot \mathbf{x}}
	\ket{ \mathbf{z} } \ .
\end{align}

Thus, after Alice receives the registers back, she applies the Hadamard transform to the input register for a second time and the state becomes

\begin{align} \label{eq:BV Phase 3}
	\ket{\psi_3}
	&= \frac{1}{\sqrt{2^n}}
	\sum_{\mathbf{x} \in \{ 0, 1 \}^n}
	(-1)^{\mathbf{s} \cdot \mathbf{x}}
	H^{\otimes n} \ket{\mathbf{x}}
	\left( \frac{\ket{0} - \ket{1}}{\sqrt{2}} \right)
	\overset{ (\ref{eq:Walsh-Hadamard Transform on Arbitrary Basis Ket $x$}) } { = }
	\frac{1}{\sqrt{2^n}}
	\sum_{ \mathbf{x} \in \{ 0, 1 \}^n }^{}
	(-1)^{\mathbf{s} \cdot \mathbf{x}}
	\left(
	\frac{1}{\sqrt{2^n}}
	\sum_{ \mathbf{z} \in \{ 0, 1 \}^n }^{}
	(-1)^{\mathbf{z} \cdot \mathbf{x}}
	\ket{\mathbf{z}}
	\right)
	\left( \frac{\ket{0} - \ket{1}}{\sqrt{2}} \right)
	\nonumber \\
	&=
	\frac{1}{2^n}
	\sum_{ \mathbf{x} \in \{ 0, 1 \}^n }^{}
	\sum_{ \mathbf{z} \in \{ 0, 1 \}^n }^{}
	(-1)^{\mathbf{s} \cdot \mathbf{x} \oplus \mathbf{z} \cdot \mathbf{x}}
	\ket{\mathbf{z}}
	\left( \frac{\ket{0} - \ket{1}}{\sqrt{2}} \right)
	=
	\frac{1}{2^n}
	\sum_{ \mathbf{z} \in \{ 0, 1 \}^n }^{}
	\sum_{ \mathbf{x} \in \{ 0, 1 \}^n }^{}
	(-1)^{ ( \mathbf{s} \oplus \mathbf{z} ) \cdot \mathbf{x} }
	\ket{\mathbf{z}}
	\left( \frac{\ket{0} - \ket{1}}{\sqrt{2}} \right)
	\nonumber \\
	&=
	\ket{\mathbf{s}}
	\left( \frac{\ket{0} - \ket{1}}{\sqrt{2}} \right)
\end{align}

The last equation is due to the following fact: if $\mathbf{s} = \mathbf{z}$, then $\forall \ \mathbf{x} \in \{ 0, 1 \}^n \ ( \mathbf{s} \oplus \mathbf{z} ) \cdot \mathbf{x} = 0$, otherwise for exactly half of the inputs $\mathbf{x}$ the exponent will be $0$ and for the remaining half the exponent will be $1$. This is typically written in a more concise manner as follows:

\begin{align} \label{eq:Inner Product Modulo 2 Property}
	\sum_{\mathbf{x} \in \{ 0, 1 \}^n }^{}
	(-1)^{ ( \mathbf{s} \oplus \mathbf{z} ) \cdot \mathbf{x} }
	=
	2^{n} \delta_{\mathbf{s}, \mathbf{z}} \ .
\end{align}

The algorithm terminates with the final measurement of the input register by Alice whereby she obtains the secret key $\mathbf{s}$ and concludes the whole process.

\section{QKD based on symmetric entangled B-V} \label{sec:QKD based on Symmetric Entangled B-V}

In this section the two versions of the proposed symmetric entangled QKD protocol based on the Bernstein-Vazirani algorithm are presented and described in great detail. These are the \emph{fully symmetric} version of the protocol, or \textbf{fSEBV} for short, and the \emph{semi-symmetric} version of the protocol, or \textbf{sSEBV} for short.

\subsection{The fSEBV protocol}

Starting with the fSEBV protocol we consider a slight alteration of the aforementioned Bernstein-Vazirani game. As before, the game starts with the two players Alice and Bob who are spatially separated. This time, instead of using normal qubits in a separable state, they use maximally entangled EPR pairs, and they both share a qubit from each pair. An important  rule of the game is that there are no limitations, on which individual will actually create the EPR pairs in the first place. The pairs can be created and distributed accordingly by Alice or Bob, or they can be acquired from a third party source. This last situation is depicted in Figure \ref{fig:Alice, Bob and Source of Entangled Pairs}. Exactly as in the previous game, the goal of the current game is to acquire a secret key $\mathbf{s}$. However, in this specific protocol symmetry plays a crucial role, as Alice and Bod behave in a perfectly symmetrical way, by both having their own secret keys, which they will attempt to input into the system, exactly as in the original algorithm. Alice's key is denoted by $\mathbf{s}_A$, Bob's key by $\mathbf{s}_B$ and they both take identical actions. Note that neither Alice, nor Bob need apply the Hadamard transform onto their input registers because they are already in the desired even superposition of all basis states, as they are populated by $n$ pairs in the $\ket{\Phi^{+}}$ Bell state. In this respect the fSEBV protocol differs from the vanilla Bernstein-Vazirani algorithm.

\begin{figure}[H]
	\centering
	\begin{tikzpicture} [scale = 1.0]
		\scoped [on background layer]
		\fill [ yellow, line width = 1.5pt, rounded corners = 12pt](3.0, 0.0) rectangle (4.0, 3.0);
		\node (Alice) at (3.5, -0.5){\textbf{Source of $\ket{\Phi^{+}}$ pairs}};
		\scoped [on background layer]
		\fill [ RedPurple!15, line width = 1.5pt, rounded corners = 12pt](-4.0, 0.0) rectangle (-1.0, 3.0);
		\node (Alice) at (-2.5, 1.5){\textbf{Alice}};
		\scoped [on background layer]
		\fill [GreenLighter2!20, line width = 1.5pt, rounded corners = 12pt](8.0, 0.0) rectangle (11.0, 3.0);
		\node (Bob) at (9.5, 1.5){\textbf{Bob}};
		\draw [WordTurquoiseLighter80, <->, >=stealth, line width = 7.0 mm] (-1.0, 1.5) -- (8.0, 1.5);
		\node () at (3.5, 1.5){ Quantum Channel };
		\draw [WordIceBlue, -, >=stealth, line width = 7.0 mm] (-2.75, 5.0) -- (9.75, 5.0);
		\node () at (3.5, 5.0){ Public Channel };
		\draw [WordIceBlue, ->, >=stealth, line width = 5.0 mm] (-2.5, 5.25) -- (-2.5, 3.0);
		\draw [WordIceBlue, ->, >=stealth, line width = 5.0 mm] (9.5, 5.25) -- (9.5, 3.0);
		\node [circle, fill = WordAquaDarker50, minimum size = 0.5 mm] () at (0.0, 1.5) {  };
		\node () at (0.0, 1.0) { $\ket{q_0}_{A}$  };
		\node [circle, fill = WordAquaDarker50, minimum size = 0.5 mm] () at (7.0, 1.5) {  };
		\node () at (7.0, 1.0) { $\ket{q_0}_{B}$ };
		\node () at (0.75, 1.5) { $\dots$ };
		\node () at (6.25, 1.5) { $\dots$ };
		\node [circle, fill = RedPurple, minimum size = 0.5 mm] () at (1.5, 1.5) {  };
		\node () at (1.5, 1.0) { $\ket{q_{n - 1}}_{A}$  };
		\node [circle, fill = RedPurple, minimum size = 0.5 mm] () at (5.5, 1.5) {  };
		\node () at (5.5, 1.0) { $\ket{q_{n - 1}}_{B}$  };
	\end{tikzpicture}
	\caption{Alice and Bob are spatially separated. A third party, the source, creates $n$ pairs of $\ket{\Phi^{+}}$ entangled photons and sends one qubit from every pair to Alice and the other qubit to Bob.}
	\label{fig:Alice, Bob and Source of Entangled Pairs}
\end{figure}
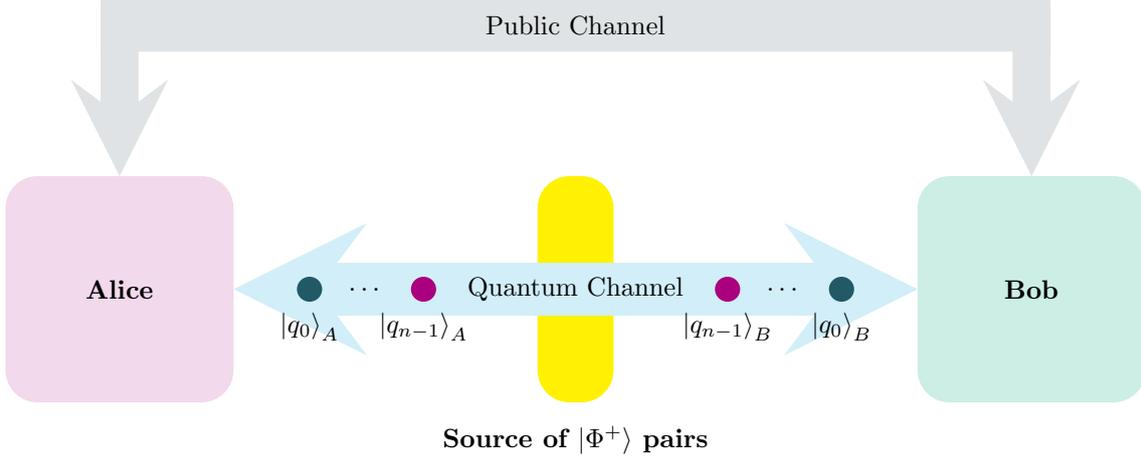

Following the aforementioned steps of the fSEBV protocol, a valid question may arise regarding what will Alice and Bob acquire after they both apply their starting secret keys $\mathbf{s}_{A}$ and $\mathbf{s}_{B}$ into their own pieces of the EPR pairs? In order to provide the answer, let us examine the algorithm more closely, by starting with the initial state of the system

\begin{align} \label{eq:QKD BV Intial State}
	\ket{\psi_0}
	=
	\ket{\Phi^{+}}^{\otimes n}
	\ket{1}_{A} \ket{1}_{B}
	\overset{ (\ref{eq:N Bell States Phi +}) } { = }
	\frac{1}{ \sqrt{2^n} }
	\sum_{\mathbf{x} \in \{ 0, 1 \}^n}
	\ket{\mathbf{x}}_{A} \ket{\mathbf{x}}_{B}
	\ket{1}_{A} \ket{1}_{B}
	\ .
\end{align}

Subscripts A and B are consistently used to designate the registers of Alice and Bob, respectively. Alice and Bob initiate the protocol by applying the Hadamard transform to their output registers, which produces the ensuing state

\begin{align} \label{eq:QKD BV Phase 1}
	\ket{\psi_1}
	=
	\frac{1}{ \sqrt{2^n} }
	\sum_{\mathbf{x} \in \{ 0, 1 \}^n}
	\ket{\mathbf{x}}_{A} \ket{\mathbf{x}}_{B}
	\left( \frac{\ket{0} - \ket{1}}{\sqrt{2}} \right)_{A} \left( \frac{\ket{0} - \ket{1}}{\sqrt{2}} \right)_{B} \ .
\end{align}

Now, both Alice and Bob can apply their functions on their registers by using the standard scheme

\begin{align} \label{eq:QKD BV Oracle}
	U_{f} : \ket{\mathbf{x}, y} \rightarrow \ket{\mathbf{x}, y\oplus f(\mathbf{x})} \ .
\end{align}

Consequently, the next state becomes

\begin{align} \label{eq:QKD BV Phase 2 - I}
	\ket{\psi_2}
	=
	\frac{1}{ \sqrt{2^n} }
	\sum_{\mathbf{x} \in \{ 0, 1 \}^n}
	(-1)^{f_{A}(\mathbf{x})} \ket{\mathbf{x}}_{A} (-1)^{f_{B}(\mathbf{x})} \ket{\mathbf{x}}_{B}
	\left( \frac{\ket{0} - \ket{1}}{\sqrt{2}} \right)_{A} \left( \frac{\ket{0} - \ket{1}}{\sqrt{2}} \right)_{B} \ .
\end{align}

At this stage let us recall that Alice's and Bob's functions are

\begin{minipage}[b]{0.45 \textwidth}
	\begin{align} \label{eq:Alice's Function}
		f_{ A } ( \mathbf{x} ) = \mathbf{s}_{A} \cdot \mathbf{x} \bmod 2
	\end{align}
\end{minipage} 
\hfill
\begin{minipage}[b]{0.45 \textwidth}
	\begin{align} \label{eq:Bob's Function}
		f_{ B } ( \mathbf{x} ) = \mathbf{s}_{B} \cdot \mathbf{x} \bmod 2 \ ,
	\end{align}
\end{minipage} 

where $\mathbf{s}_{A}$ and $\mathbf{s}_{B}$ are the keys chosen by Alice and Bob, respectively. Based on (\ref{eq:Alice's Function}) and (\ref{eq:Bob's Function}), (\ref{eq:QKD BV Phase 2 - I}) can be written as

\begin{align} \label{eq:QKD BV Phase 2 - II}
	\ket{\psi_2}
	&=
	\frac{1}{ \sqrt{2^n} }
	\sum_{\mathbf{x} \in \{ 0, 1 \}^n}
	(-1)^{\mathbf{s}_{A} \cdot \mathbf{x}} \ket{\mathbf{x}}_{A} (-1)^{\mathbf{s}_{B} \cdot \mathbf{x}} \ket{\mathbf{x}}_{B}
	\left( \frac{\ket{0} - \ket{1}}{\sqrt{2}} \right)_{A} \left( \frac{\ket{0} - \ket{1}}{\sqrt{2}} \right)_{B}
	\nonumber \\
	&=
	\frac{1}{ \sqrt{2^n} }
	\sum_{\mathbf{x} \in \{ 0, 1 \}^n}
	(-1)^{\mathbf{s}_{A} \cdot \mathbf{x} \oplus \mathbf{s}_{B} \cdot \mathbf{x}}
	\ket{\mathbf{x}}_{A} \ket{\mathbf{x}}_{B}
	\left( \frac{\ket{0} - \ket{1}}{\sqrt{2}} \right)_{A} \left( \frac{\ket{0} - \ket{1}}{\sqrt{2}} \right)_{B}
	\nonumber \\
	&=
	\frac{1}{ \sqrt{2^n} }
	\sum_{\mathbf{x} \in \{ 0, 1 \}^n}
	(-1)^{ ( \mathbf{s}_{A} \oplus \mathbf{s}_{B} ) \cdot \mathbf{x}}
	\ket{\mathbf{x}}_{A} \ket{\mathbf{x}}_{B}
	\left( \frac{\ket{0} - \ket{1}}{\sqrt{2}} \right)_{A} \left( \frac{\ket{0} - \ket{1}}{\sqrt{2}} \right)_{B}
	\ .
\end{align}

Subsequently, both Alice and Bob apply the Hadamard transformation to their input registers, driving the system into the next state

\begin{align} \label{eq:QKD BV Phase 3 - I}
	\ket{\psi_3}
	&=
	\frac{1}{ \sqrt{2^n} }
	\sum_{\mathbf{x} \in \{ 0, 1 \}^n}
	(-1)^{ ( \mathbf{s}_{A} \oplus \mathbf{s}_{B} ) \cdot \mathbf{x} }
	H^{\otimes n} \ket{\mathbf{x}}_{A} H^{\otimes n} \ket{\mathbf{x}}_{B}
	\left( \frac{\ket{0} - \ket{1}}{\sqrt{2}} \right)_{A} \left( \frac{\ket{0} - \ket{1}}{\sqrt{2}} \right)_{B}
	\nonumber \\
	&\overset{ (\ref{eq:Walsh-Hadamard Transform on Arbitrary Basis Ket $x$}) } { = }
	\frac{1}{ \sqrt{2^n} }
	\sum_{ \mathbf{x} \in \{ 0, 1 \}^n }^{}
	(-1)^{ ( \mathbf{s}_{A} \oplus \mathbf{s}_{B} ) \cdot \mathbf{x} }
	\left(
	\frac{1}{\sqrt{2^n}}
	\sum_{ \mathbf{z} \in \{ 0, 1 \}^n }^{}
	(-1)^{\mathbf{z} \cdot \mathbf{x}}
	\ket{\mathbf{z}}_{A}
	\right)
	\left(
	\frac{1}{\sqrt{2^n}}
	\sum_{ \mathbf{w} \in \{ 0, 1 \}^n }^{}
	(-1)^{\mathbf{w} \cdot \mathbf{x}}
	\ket{\mathbf{w}}_{B}
	\right)
	\nonumber \\
	&\left( \frac{\ket{0} - \ket{1}}{\sqrt{2}} \right)_{A}
	\left( \frac{\ket{0} - \ket{1}}{\sqrt{2}} \right)_{B}
	\nonumber \\
	&=
	\frac{1}{ ( \sqrt{2^n} )^{3} }
	\sum_{ \mathbf{x} \in \{ 0, 1 \}^n }^{}
	\sum_{ \mathbf{z} \in \{ 0, 1 \}^n }^{}
	\sum_{ \mathbf{w} \in \{ 0, 1 \}^n }^{}
	(-1)^{ ( \mathbf{s}_{A} \oplus \mathbf{s}_{B} \oplus \mathbf{z} \oplus \mathbf{w} ) \cdot \mathbf{x} }
	\ket{\mathbf{z}}_{A} \ket{\mathbf{w}}_{B}
	\left( \frac{\ket{0} - \ket{1}}{\sqrt{2}} \right)_{A} \left( \frac{\ket{0} - \ket{1}}{\sqrt{2}} \right)_{B}
	\nonumber \\
	&=
	\frac{1}{ ( \sqrt{2^n} )^{3} }
	\sum_{ \mathbf{z} \in \{ 0, 1 \}^n }^{}
	\sum_{ \mathbf{w} \in \{ 0, 1 \}^n }^{}
	\sum_{ \mathbf{x} \in \{ 0, 1 \}^n }^{}
	(-1)^{ ( \mathbf{s}_{A} \oplus \mathbf{s}_{B} \oplus \mathbf{z} \oplus \mathbf{w} ) \cdot \mathbf{x} }
	\ket{\mathbf{z}}_{A} \ket{\mathbf{w}}_{B}
	\left( \frac{\ket{0} - \ket{1}}{\sqrt{2}} \right)_{A} \left( \frac{\ket{0} - \ket{1}}{\sqrt{2}} \right)_{B}
	\ .
\end{align}

When $\mathbf{z} \oplus \mathbf{w} = \mathbf{s}_{A} \oplus \mathbf{s}_{B}$, then $\forall  \mathbf{x} \in \{ 0, 1 \}^n$, the expression $(-1)^{ ( \mathbf{s}_{A} \oplus \mathbf{s}_{B} \oplus \mathbf{z} \oplus \mathbf{w} ) \cdot \mathbf{x} }$ becomes $(-1)^{0} = 1$ and the sum $\sum_{ \mathbf{x} \in \{ 0, 1 \}^n }^{}
(-1)^{ ( \mathbf{s}_{A} \oplus \mathbf{s}_{B} \oplus \mathbf{z} \oplus \mathbf{w} ) \cdot \mathbf{x} } = 2^{n}$. Whenever $\mathbf{z} \oplus \mathbf{w} \neq \mathbf{s}_{A} \oplus \mathbf{s}_{B}$, the sum is just $0$ because for exactly half of the inputs $\mathbf{x}$ the exponent will be $0$ and for the remaining half the exponent will be $1$. Hence, one may write that

\begin{align} \label{eq:Entangled Inner Product Modulo 2 Property}
	\sum_{\mathbf{x} \in \{ 0, 1 \}^n }^{}
	(-1)^{ ( \mathbf{s}_{A} \oplus \mathbf{s}_{B} \oplus \mathbf{z} \oplus \mathbf{w} ) \cdot \mathbf{x} }
	=
	2^{n} \delta_{\mathbf{s}_{A} \oplus \mathbf{s}_{B}, \mathbf{z} \oplus \mathbf{w}}
	\ .
\end{align}

Utilizing equation (\ref{eq:Entangled Inner Product Modulo 2 Property}), and ignoring for the moment the two factors $\left( \frac{\ket{0} - \ket{1}}{\sqrt{2}} \right)_{A}$ and $\left( \frac{\ket{0} - \ket{1}}{\sqrt{2}} \right)_{B}$, the following two equivalent and symmetric forms can be derived

\begin{align} \label{eq:Entangled Triple Sum Reduction - I}
	& \sum_{ \mathbf{z} \in \{ 0, 1 \}^n }^{}
	\sum_{ \mathbf{w} \in \{ 0, 1 \}^n }^{}
	\sum_{ \mathbf{x} \in \{ 0, 1 \}^n }^{}
	(-1)^{ ( \mathbf{s}_{A} \oplus \mathbf{s}_{B} \oplus \mathbf{z} \oplus \mathbf{w} ) \cdot \mathbf{x} }
	\ket{\mathbf{z}}_{A}
	\ket{\mathbf{w}}_{B}
	=
	2^{n}
	\sum_{ \mathbf{z} \in \{ 0, 1 \}^n }^{}
	\ket{ \mathbf{z} }_{A}
	\ket{ \mathbf{s}_{A} \oplus \mathbf{s}_{B} \oplus \mathbf{z} }_{B}
	\ ,
\end{align}

and

\begin{align} \label{eq:Entangled Triple Sum Reduction - II}
	& \sum_{ \mathbf{w} \in \{ 0, 1 \}^n }^{}
	\sum_{ \mathbf{z} \in \{ 0, 1 \}^n }^{}
	\sum_{ \mathbf{x} \in \{ 0, 1 \}^n }^{}
	(-1)^{ ( \mathbf{s}_{A} \oplus \mathbf{s}_{B} \oplus \mathbf{z} \oplus \mathbf{w} ) \cdot \mathbf{x} }
	\ket{\mathbf{z}}_{A}
	\ket{\mathbf{w}}_{B}
	=
	2^{n}
	\sum_{ \mathbf{w} \in \{ 0, 1 \}^n }^{}
	\ket{ \mathbf{s}_{A} \oplus \mathbf{s}_{B} \oplus \mathbf{w} }_{A}
	\ket{ \mathbf{w} }_{B}
	\ .
\end{align}

By combining (\ref{eq:QKD BV Phase 3 - I}) with (\ref{eq:Entangled Triple Sum Reduction - I}) and (\ref{eq:Entangled Triple Sum Reduction - II}), state $\ket{\psi_3}$ can be written in two different ways:

\begin{align} \label{eq:QKD BV Phase 3 - II}
	\ket{\psi_3}
	&=
	\frac{1}{ \sqrt{2^n} }
	\sum_{ \mathbf{z} \in \{ 0, 1 \}^n }^{}
	\ket{ \mathbf{z} }_{A}
	\ket{ \mathbf{s}_{A} \oplus \mathbf{s}_{B} \oplus \mathbf{z} }_{B}
	\left( \frac{\ket{0} - \ket{1}}{\sqrt{2}} \right)_{A} \left( \frac{\ket{0} - \ket{1}}{\sqrt{2}} \right)_{B}
	\nonumber \\
	&=
	\frac{1}{ \sqrt{2^n} }
	\sum_{ \mathbf{w} \in \{ 0, 1 \}^n }^{}
	\ket{ \mathbf{s}_{A} \oplus \mathbf{s}_{B} \oplus \mathbf{w} }_{A}
	\ket{ \mathbf{w} }_{B}
	\left( \frac{\ket{0} - \ket{1}}{\sqrt{2}} \right)_{A} \left( \frac{\ket{0} - \ket{1}}{\sqrt{2}} \right)_{B}
	\ .
\end{align}

Finally, Alice and Bob measure their EPR pairs in the input registers, getting

\begin{align} \label{eq:QKD BV Final Measurement}
	\ket{\psi_4}
	=
	\ket{ \mathbf{z}_{0} }_{A}
	\ket{ \mathbf{s}_{A} \oplus \mathbf{s}_{B} \oplus \mathbf{z}_{0} }_{B}
	=
	\ket{ \mathbf{s}_{A} \oplus \mathbf{s}_{B} \oplus \mathbf{w}_{0} }_{A}
	\ket{ \mathbf{w}_{0} }_{B}
	\ ,
	\quad \text{for some} \quad
	\mathbf{z}_{0}, \mathbf{w}_{0} \in \{ 0, 1 \}^n
	\ .
\end{align}

Note that in general $\mathbf{z}_{0} \neq \mathbf{w}_{0}$. The quantum part of the protocol is now complete. The final secret key is the string $\mathbf{s}_{A} \oplus \mathbf{s}_{B} \oplus \mathbf{z}_{0}$ that Bob measured in his input register. In the highly unlikely event that $\ket { \mathbf{s}_{A} \oplus \mathbf{s}_{B} \oplus \mathbf{z}_{0} } = \ket{0}^{\otimes n}$, Bob should inform Alice through the use of the public channel that the whole procedure must be repeated once again, since such a key is clearly unacceptable. However, for a $n$-bit key the probability of this happening is negligible, specifically $\frac{1}{2^n}$, which rapidly tends to $0$ as $n \to \infty$. Hence, it may be safely assumed that Bob possesses a viable secret key, namely $\mathbf{s}_{A} \oplus \mathbf{s}_{B} \oplus \mathbf{z}_{0}$. Now the final step is for Alice to obtain the secret key too. This is easily achieved by simply having Bob publicly announce his tentative secret key $\mathbf{s}_{B}$ to Alice via the use of the public channel. Alice, who has measured the binary string $\mathbf{z}_{0}$ and she is already aware of her initial secret key $\mathbf{s}_{A}$, can easily obtain the final key, by simply calculating the XOR of $\mathbf{s}_{A}$, her measurement $\mathbf{z}_{0}$ and Bob's initial key $\mathbf{s}_{B}$, which she learns from the public channel. This concludes the fSEBV protocol.

The symmetry inherent in this protocol, enables the seamless reversal of roles. Figure \ref{fig:Alice's and Bob's Actions in the fSEBV Protocol} grants the initiative to Bob: it is his measurement $\mathbf{s}_{A} \oplus \mathbf{s}_{B} \oplus \mathbf{z}_{0}$ that produces the secret key and it is his task to send his initial key $\mathbf{s}_{B}$ to Alice, in order to successfully complete the procedure. It is eqaually feasible to have Alice instead of Bob drive the whole process by taking her measurement $\mathbf{s}_{A} \oplus \mathbf{s}_{B} \oplus \mathbf{w}_{0}$ to be the secret key, as shown in (\ref{eq:QKD BV Final Measurement}). In such an implementation of the fSEBV protocol, Alice must reveal her initial key $\mathbf{s}_{A}$ to Bob via the public channel.

During the transmission of Bob's key $\mathbf{s}_{B}$ using a public channel, any potential eavesdropper, namely Eve, does not gain any advantage by listening to the public channel. Due to the fact that she is oblivious of $\mathbf{z}_{0}$ and $\mathbf{s}_{A}$. Thus, she has no way of knowing or computing the final secret key. Hence, the fSEBV protocol ensures that if Alice and Bob can create their keys using a random number generator, in order to avoid possible patterns in the keys, Eve will be left with $2^n$ different combinations to test in order to find the secret key.

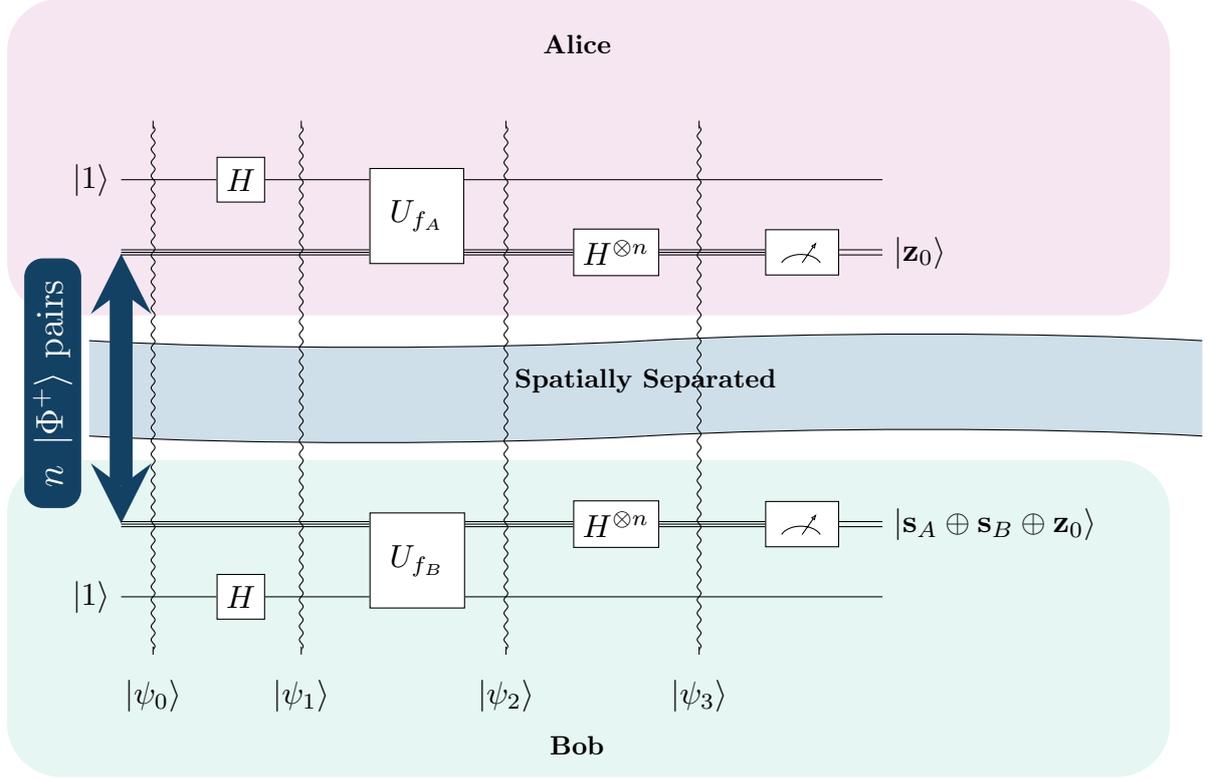
\begin{figure}[H]
	\begin{center}
		\begin{tikzpicture} [scale = 1.2]
			\scoped [on background layer]
			\fill [RedPurple!10, line width = 1.5pt, rounded corners = 20pt](-1.25, -2.25) rectangle (11.5, 1.25);
			\node (Alice) at (5.0, 0.75){\textbf{Alice}};
			\scoped [on background layer]
			\fill [GreenLighter2!10, line width = 1.5pt, rounded corners = 20pt](-1.25, -7.35) rectangle (11.5, -3.85);
			\node (Bob) at (5.0, -7.0){\textbf{Bob}};
			\draw [WordBlue, <->, >=stealth, line width = 3.0 mm] (-0.0, -4.55) -- (-0.0, -1.57);
			\node [ rotate = 90, rectangle, rounded corners = 7pt, fill = WordBlue, text = white ] at (-0.75, -3.0) { \Large $ \ n \ \ket{\Phi^{+}}$ pairs \ };
			\begin{yquant} 
				nobit AAUX1;
				qubit {$\ket{1}$} AOR;
				qubits {} AIR;
				nobit AAUX2;
				nobit AAUX3;
				[name = wave, register/minimum height = 5mm, register/minimum depth = 5mm]
				nobit wave;
				nobit BAUX3;
				nobit BAUX2;
				qubits {} BIR;
				qubit {$\ket{1}$} BOR;
				nobit BAUX1;
				[ name = S0 ]
				barrier (AAUX1, BAUX1);
				hspace {0.3 cm} AOR;
				align AIR, AOR, BOR, BIR;
				[ fill = white ]
				[ fill = white ]
				box {$H$} AOR;
				[ fill = white ]
				box {$H$} BOR;
				[ fill = white ]
				%
				[ name = S1 ]
				barrier (AAUX1, BAUX1);
				hspace {0.3 cm} AIR;
				align AIR, BIR;
				[ fill = white ]
				box {$ \ U_{f_{A}} \ $} (AOR, AIR);
				[ fill = white ]
				box {$ \ U_{f_{B}} \ $} (BOR, BIR);
				[ name = S2 ]
				barrier (AAUX1, BAUX1);
				hspace {0.3 cm} AIR;
				align AIR, BIR;
				[ fill = white ]
				box {$H^{\otimes n}$} AIR;
				[ fill = white ]
				box {$H^{\otimes n}$} BIR;
				[ name = S3 ]
				barrier (AAUX1, BAUX1);
				hspace {0.3 cm} AIR;
				[ fill = white ]
				measure AIR;
				hspace {0.3 cm} BIR;
				[ fill = white ]
				measure BIR;
				hspace {0.3 cm} AIR;
				output {$\ket{ \mathbf{z}_{0} }$} AIR;
				hspace {0.3 cm} BIR;
				output {$\ket{ \mathbf{s}_{A} \oplus \mathbf{s}_{B} \oplus \mathbf{z}_{0} }$} BIR;
				\scoped [on background layer]
				\node [wave, fill = WordDarkTealLighter80, fit = (wave) (current bounding box.east |- wave), inner ysep = 15.0pt, inner xsep = 12pt] {\textbf{Spatially Separated}};
				\node [ below = 6.0 cm ] at (S0) {$\ket{\psi_{0}}$};
				\node [ below = 6.0 cm ] at (S1) {$\ket{\psi_{1}}$};
				\node [ below = 6.0 cm ] at (S2) {$\ket{\psi_{2}}$};
				\node [ below = 6.0 cm ] at (S3) {$\ket{\psi_{3}}$};
			\end{yquant}
		\end{tikzpicture}
	\end{center}
	\caption{This figures gives a schematic representation of the proposed protocol.}
	\label{fig:Symmetric Entangled B-V Protocol}
\end{figure}

Figure \ref{fig:Alice's and Bob's Actions in the fSEBV Protocol} summarizes the steps of the protocol from Alice's and Bob's side in an algorithmic manner. Figure \ref{fig:Symmetric Entangled B-V Protocol} depicts the protocol graphically in the form of a quantum circuit.

\SetAlgorithmName{Protocol fSEBV}{ }{ }
%
\begin{figure}[H]
	\begin{minipage}[t]{7.0 cm}
		\begin{algorithm}[H] 
			\caption{ Alice's actions }
			\label{alg: Alice's Actions in fSEBV Protocol}
			\vspace{0.3 cm}
			$\bullet$ Alice's input register is populated with entangled qubits
			\\
			$\bullet$ Alice's output register is set to $\ket 1$
			\\
			$\bullet$ Alice applies the Hadamard transform to her output register
			\\
			$\bullet$ Alice applies her tentative key $\mathbf{s}_A$
			\\
			$\bullet$ Alice applies the Hadamard transform to her input register
			\\
			$\bullet$ Alice measures her input register to find the random binary string $\mathbf{z}_0$
			\\
			$\bullet$ Alice receives information from Bob whether the process was a success or must be repeated
			\\
			$\bullet$ If the procedure was successful, Alice receives from Bob his key $\mathbf{s}_B$ and, by already knowing $\mathbf{s}_A$ and $\mathbf{z}_0$, she computes the final key $\mathbf{s}_{A} \oplus \mathbf{s}_{B} \oplus \mathbf{z}_{0}$
			\\
		\end{algorithm}
	\end{minipage}
	\begin{minipage}[t]{7.0 cm}
		\begin{algorithm}[H] 
			\caption{ Bob's actions }
			\label{alg: Bob's Actions in fSEBV Protocol}
			\vspace{0.3 cm}
			$\bullet$ Bob's input register is populated with entangled qubits
			\\
			$\bullet$ Bob's output register is set to $\ket 1$
			\\
			$\bullet$ Bob applies the Hadamard transform to his output register
			\\
			$\bullet$ Bob applies his tentative key $\mathbf{s}_B$
			\\
			$\bullet$ Bob applies the Hadamard transform to his input register
			\\
			$\bullet$ Bob measures his input register to find the final secret key $\mathbf{s}_{A} \oplus \mathbf{s}_{B} \oplus \mathbf{z}_{0}$
			\\
			$\bullet$ In the unlikely event that $\ket { \mathbf{s}_{A} \oplus \mathbf{s}_{B} \oplus \mathbf{z}_{0} } = \ket{0}^{\otimes n}$, Bob informs Alice that the process must be repeated from the start
			\\
			$\bullet$ Otherwise Bob communicates his tentative key $\mathbf{s}_B$ to Alice via the public channel
		\end{algorithm}
	\end{minipage}
	\caption{Alice's and Bob's actions during the application of the fSEBV protocol.}
	\label{fig:Alice's and Bob's Actions in the fSEBV Protocol}
\end{figure}

\vspace{0.3 cm}

\subsection{The sSEBV protocol}

The sSEBV protocol explores a special but important case of the sSEBV protocol, which differs from the latter in one important aspect. Alice possess her random initial key $\mathbf{s}_{A}$, but Bob's key $\mathbf{s}_{B}$ is not a random binary string anymore; it is specifically taken to be $\mathbf{0} = 0 \dots 0$. Essentially, sSEBV protocol answers the question of what will happen, if one of the players, either Alice or Bob, decides not to send a key. As before Alice and Bob are spatially separated and they both share $n$ EPR pairs. In this variant, Alice and Bod behave in a semi-symmetrical way. Alice possess her random initial key $\mathbf{s}_A$, but Bob is obliged to use $\mathbf{0}$ as his initial key.

In this case too the initial state of the system is

\begin{align} \label{eq:QKD BV Intial State II}
	\ket{\psi_0}
	=
	\ket{\Phi^{+}}^{\otimes n}
	\ket{1}_{A} \ket{1}_{B}
	\overset{ (\ref{eq:N Bell States Phi +}) } { = }
	\frac{1}{ \sqrt{2^n} }
	\sum_{\mathbf{x} \in \{ 0, 1 \}^n}
	\ket{\mathbf{x}}_{A} \ket{\mathbf{x}}_{B}
	\ket{1}_{A} \ket{1}_{B}
	\ .
\end{align}

Similarly, Alice and Bob initiate the protocol by applying the Hadamard transform to their output registers, which produces the ensuing state

\begin{align} \label{eq:QKD BV Phase 1 - II}
	\ket{\psi_1}
	=
	\frac{1}{ \sqrt{2^n} }
	\sum_{\mathbf{x} \in \{ 0, 1 \}^n}
	\ket{\mathbf{x}}_{A} \ket{\mathbf{x}}_{B}
	\left( \frac{\ket{0} - \ket{1}}{\sqrt{2}} \right)_{A} \left( \frac{\ket{0} - \ket{1}}{\sqrt{2}} \right)_{B} \ .
\end{align}

Next Alice and Bob apply their corresponding functions on their registers via the standard scheme

\begin{align} \label{eq:QKD BV Oracle - II}
	U_{f} : \ket{\mathbf{x}, y} \rightarrow \ket{\mathbf{x}, y\oplus f(\mathbf{x})} \ ,
\end{align}

only now the situation is quite different because Bob must necessarily use $\mathbf{0}$:

\begin{minipage}[b]{0.45 \textwidth}
	\begin{align} \label{eq:Alice's Function - II}
		f_{ A } ( \mathbf{x} ) = \mathbf{s}_{A} \cdot \mathbf{x} \bmod 2
	\end{align}
\end{minipage} 
\hfill
\begin{minipage}[b]{0.45 \textwidth}
	\begin{align} \label{eq:Bob's Function - II}
		f_{ B } ( \mathbf{x} ) =  \mathbf{0} \cdot \mathbf{x} \bmod 2 = 0 \ .
	\end{align}
\end{minipage} 

Therefore, the next state becomes

\begin{align} \label{eq:QKD BV Phase 2 - I - II}
	\ket{\psi_2}
	&=
	\frac{1}{ \sqrt{2^n} }
	\sum_{\mathbf{x} \in \{ 0, 1 \}^n}
	(-1)^{f_{A}(\mathbf{x})} \ket{\mathbf{x}}_{A} (-1)^{0} \ket{\mathbf{x}}_{B}
	\left( \frac{\ket{0} - \ket{1}}{\sqrt{2}} \right)_{A} \left( \frac{\ket{0} - \ket{1}}{\sqrt{2}} \right)_{B}
	\nonumber \\
	&\overset{ (\ref{eq:Alice's Function - II}) } { = }
	\frac{1}{ \sqrt{2^n} }
	\sum_{\mathbf{x} \in \{ 0, 1 \}^n}
	(-1)^{ \mathbf{s}_{A} \cdot \mathbf{x}}
	\ket{\mathbf{x}}_{A} \ket{\mathbf{x}}_{B}
	\left( \frac{\ket{0} - \ket{1}}{\sqrt{2}} \right)_{A} \left( \frac{\ket{0} - \ket{1}}{\sqrt{2}} \right)_{B}
	 \ .
\end{align}

Subsequently, both Alice and Bob apply the Hadamard transformation to their input registers, driving the system into the next state

\begin{align} \label{eq:QKD BV Phase 3 - I - II}
	\ket{\psi_3}
	&=
	\frac{1}{ \sqrt{2^n} }
	\sum_{\mathbf{x} \in \{ 0, 1 \}^n}
	(-1)^{ \mathbf{s}_{A} \cdot \mathbf{x}}
	H^{\otimes n} \ket{\mathbf{x}}_{A} H^{\otimes n} \ket{\mathbf{x}}_{B}
	\left( \frac{\ket{0} - \ket{1}}{\sqrt{2}} \right)_{A} \left( \frac{\ket{0} - \ket{1}}{\sqrt{2}} \right)_{B}
	\nonumber \\
	&\overset{ (\ref{eq:Walsh-Hadamard Transform on Arbitrary Basis Ket $x$}) } { = }
	\frac{1}{ \sqrt{2^n} }
	\sum_{ \mathbf{x} \in \{ 0, 1 \}^n }^{}
	(-1)^{ \mathbf{s}_{A} \cdot \mathbf{x}}
	\left(
	\frac{1}{\sqrt{2^n}}
	\sum_{ \mathbf{z} \in \{ 0, 1 \}^n }^{}
	(-1)^{\mathbf{z} \cdot \mathbf{x}}
	\ket{\mathbf{z}}_{A}
	\right)
	\left(
	\frac{1}{\sqrt{2^n}}
	\sum_{ \mathbf{w} \in \{ 0, 1 \}^n }^{}
	(-1)^{\mathbf{w} \cdot \mathbf{x}}
	\ket{\mathbf{w}}_{B}
	\right)
	\nonumber \\
	&\left( \frac{\ket{0} - \ket{1}}{\sqrt{2}} \right)_{A}
	\left( \frac{\ket{0} - \ket{1}}{\sqrt{2}} \right)_{B}
	\nonumber \\
	&=
	\frac{1}{ ( \sqrt{2^n} )^{3} }
	\sum_{ \mathbf{x} \in \{ 0, 1 \}^n }^{}
	\sum_{ \mathbf{z} \in \{ 0, 1 \}^n }^{}
	\sum_{ \mathbf{w} \in \{ 0, 1 \}^n }^{}
	(-1)^{ ( \mathbf{s}_{A} \oplus \mathbf{z} \oplus \mathbf{w} ) \cdot \mathbf{x} }
	\ket{\mathbf{z}}_{A} \ket{\mathbf{w}}_{B}
	\left( \frac{\ket{0} - \ket{1}}{\sqrt{2}} \right)_{A} \left( \frac{\ket{0} - \ket{1}}{\sqrt{2}} \right)_{B}
	\nonumber \\
	&=
	\frac{1}{ ( \sqrt{2^n} )^{3} }
	\sum_{ \mathbf{z} \in \{ 0, 1 \}^n }^{}
	\sum_{ \mathbf{w} \in \{ 0, 1 \}^n }^{}
	\sum_{ \mathbf{x} \in \{ 0, 1 \}^n }^{}
	(-1)^{ ( \mathbf{s}_{A} \oplus \mathbf{z} \oplus \mathbf{w} ) \cdot \mathbf{x} }
	\ket{\mathbf{z}}_{A} \ket{\mathbf{w}}_{B}
	\left( \frac{\ket{0} - \ket{1}}{\sqrt{2}} \right)_{A} \left( \frac{\ket{0} - \ket{1}}{\sqrt{2}} \right)_{B}
	\ .
\end{align}

When $\mathbf{z} \oplus \mathbf{w} = \mathbf{s}_{A}$, then $\forall  \mathbf{x} \in \{ 0, 1 \}^n$, the expression $(-1)^{ ( \mathbf{s}_{A} \oplus \mathbf{z} \oplus \mathbf{w} ) \cdot \mathbf{x} }$ becomes $(-1)^{0} = 1$ and the sum $\sum_{ \mathbf{x} \in \{ 0, 1 \}^n }^{}
(-1)^{ ( \mathbf{s}_{A} \oplus \mathbf{z} \oplus \mathbf{w} ) \cdot \mathbf{x} } = 2^{n}$. Whenever $\mathbf{z} \oplus \mathbf{w} \neq \mathbf{s}_{A}$, the sum is just $0$ because for exactly half of the inputs $\mathbf{x}$ the exponent will be $0$ and for the remaining half the exponent will be $1$. So, again one may write that

\begin{align} \label{eq:Entangled Inner Product Modulo 2 Property - II}
	\sum_{\mathbf{x} \in \{ 0, 1 \}^n }^{}
	(-1)^{ ( \mathbf{s}_{A} \oplus \mathbf{z} \oplus \mathbf{w} ) \cdot \mathbf{x} }
	=
	2^{n} \delta_{\mathbf{s}_{A}, \mathbf{z} \oplus \mathbf{w}}
	\ .
\end{align}

Utilizing equation (\ref{eq:Entangled Inner Product Modulo 2 Property - II}), and ignoring for the moment the two factors $\left( \frac{\ket{0} - \ket{1}}{\sqrt{2}} \right)_{A}$ and $\left( \frac{\ket{0} - \ket{1}}{\sqrt{2}} \right)_{B}$, the following two equivalent and symmetric forms can be derived

\begin{align} \label{eq:Entangled Triple Sum Reduction - I - II}
	& \sum_{ \mathbf{z} \in \{ 0, 1 \}^n }^{}
	\sum_{ \mathbf{w} \in \{ 0, 1 \}^n }^{}
	\sum_{ \mathbf{x} \in \{ 0, 1 \}^n }^{}
	(-1)^{ ( \mathbf{s}_{A} \oplus \mathbf{s}_{B} \oplus \mathbf{z} \oplus \mathbf{w} ) \cdot \mathbf{x} }
	\ket{\mathbf{z}}_{A}
	\ket{\mathbf{w}}_{B}
	=
	2^{n}
	\sum_{ \mathbf{z} \in \{ 0, 1 \}^n }^{}
	\ket{ \mathbf{z} }_{A}
	\ket{ \mathbf{s}_{A} \oplus \mathbf{z} }_{B}
	\ ,
\end{align}

and

\begin{align} \label{eq:Entangled Triple Sum Reduction - II - II}
	& \sum_{ \mathbf{w} \in \{ 0, 1 \}^n }^{}
	\sum_{ \mathbf{z} \in \{ 0, 1 \}^n }^{}
	\sum_{ \mathbf{x} \in \{ 0, 1 \}^n }^{}
	(-1)^{ ( \mathbf{s}_{A} \oplus \mathbf{z} \oplus \mathbf{w} ) \cdot \mathbf{x} }
	\ket{\mathbf{z}}_{A}
	\ket{\mathbf{w}}_{B}
	=
	2^{n}
	\sum_{ \mathbf{w} \in \{ 0, 1 \}^n }^{}
	\ket{ \mathbf{s}_{A} \oplus \mathbf{w} }_{A}
	\ket{ \mathbf{w} }_{B}
	\ .
\end{align}

By combining (\ref{eq:QKD BV Phase 3 - I - II}) with (\ref{eq:Entangled Triple Sum Reduction - I - II}) and (\ref{eq:Entangled Triple Sum Reduction - II - II}), state $\ket{\psi_3}$ can be written in two different ways:

\begin{align} \label{eq:QKD BV Phase 3 - II - II}
	\ket{\psi_3}
	&=
	\frac{1}{ \sqrt{2^n} }
	\sum_{ \mathbf{z} \in \{ 0, 1 \}^n }^{}
	\ket{ \mathbf{z} }_{A}
	\ket{ \mathbf{s}_{A} \oplus \mathbf{z} }_{B}
	\left( \frac{\ket{0} - \ket{1}}{\sqrt{2}} \right)_{A} \left( \frac{\ket{0} - \ket{1}}{\sqrt{2}} \right)_{B}
	\nonumber \\
	&=
	\frac{1}{ \sqrt{2^n} }
	\sum_{ \mathbf{w} \in \{ 0, 1 \}^n }^{}
	\ket{ \mathbf{s}_{A} \oplus \mathbf{w} }_{A}
	\ket{ \mathbf{w} }_{B}
	\left( \frac{\ket{0} - \ket{1}}{\sqrt{2}} \right)_{A} \left( \frac{\ket{0} - \ket{1}}{\sqrt{2}} \right)_{B}
	\ .
\end{align}

Now, when Alice and Bob measure their input registers, they will get

\begin{align} \label{eq:QKD BV Final Measurement - II - II}
	\ket{\psi_4}
	=
	\ket{ \mathbf{z}_{0} }_{A}
	\ket{ \mathbf{s}_{A} \oplus \mathbf{z}_{0} }_{B}
	=
	\ket{ \mathbf{s}_{A} \oplus \mathbf{w}_{0} }_{A}
	\ket{ \mathbf{w}_{0} }_{B}
	\ ,
	\quad \text{for some} \quad
	\mathbf{z}_{0}, \mathbf{w}_{0} \in \{ 0, 1 \}^n
	\ .
\end{align}

As in the fSEBV protocol, here also holds that $\mathbf{z}_{0} \neq \mathbf{w}_{0}$ in general. This time, there are two ways in which the final part of the protocol can unfold. One way, exactly like before, is to take Bob's measurement as the new secret key. The other, equally viable choice, is to take Alice's initial key $\mathbf{s}_{A}$ as the final secret key. In that case Alice must publicly announce $\mathbf{z}_{0}$ to Bob via a public channel, so that he can compute $\mathbf{s}_{A}$. This is a suitable choice in cases where, for whatever reason, Alice must set the secret key herself, not wanting to leave anything to chance. In that way she may securely communicate her chosen key to Bob. As before, during the transmission of Alice's measurement $\mathbf{z}_{0}$ using a public channel, Eve does not gain any advantage by eavesdropping on their communication. Due to the fact that she is oblivious to $\mathbf{s}_{A}$, she has no way of knowing or computing the final secret key. Hence, the sSEBV protocol also ensures that if Alice devises her key using a random number generator, in order to avoid possible patterns in the keys, Eve will be left with $2^n$ different combinations to test in order to find the secret key.

Figure \ref{fig:Alice's and Bob's Actions in the sSEBV Protocol} presents the detailed actions for the implementation of the sSEBV protocol from Alice's and Bob's side in an algorithmic manner. Although the sSEBV protocol is not perfectly symmetric, reversal of Alice's and Bob's roles is still trivially easy. As can be seen in the description of Figure \ref{fig:Alice's and Bob's Actions in the sSEBV Protocol}, Alice is the one to choose the secret key and it is her that sends the final measurement $\mathbf{z}_{0}$ to Bob so that he can successfully derive the secret key. It is eqaually feasible to have Bob instead of Alice choose the secret key and have Alice use $\mathbf{0}$ as her key. In such a realization of the sSEBV protocol, Bob must also reveal his final measurement $\mathbf{w}_{0}$ to Alice via the public channel.

\SetAlgorithmName{Protocol sSEBV}{ }{ }
%
\begin{figure}[H]
	\begin{minipage}[t]{7.0 cm}
		\begin{algorithm}[H] 
			\caption{ Alice's actions }
			\label{alg: Alice's Actions in sSEBV Protocol}
			\vspace{0.3 cm}
			$\bullet$ Alice's input register is populated with entangled qubits
			\\
			$\bullet$ Alice's output register is set to $\ket 1$
			\\
			$\bullet$ Alice applies the Hadamard transform to her output register
			\\
			$\bullet$ Alice applies her chosen key $\mathbf{s}_A$
			\\
			$\bullet$ Alice applies the Hadamard transform to her input register
			\\
			$\bullet$ Alice measures her input register to find the random binary string $\mathbf{z}_0$
			\\
			$\bullet$ Alice announces the binary string $\mathbf{z}_0$ to Bob via the public channel
			\\
		\end{algorithm}
	\end{minipage}
	\begin{minipage}[t]{7.0 cm}
		\begin{algorithm}[H] 
			\caption{ Bob's actions }
			\label{alg: Bob's Actions in sSEBV Protocol}
			\vspace{0.3 cm}
			$\bullet$ Bob's input register is populated with entangled qubits
			\\
			$\bullet$ Bob's output register is set to $\ket 1$
			\\
			$\bullet$ Bob applies the Hadamard transform to his output register
			\\
			$\bullet$ Bob applies his key $\mathbf{0}$
			\\
			$\bullet$ Bob applies the Hadamard transform to his input register
			\\
			$\bullet$ Bob measures his input register to find the binary string $\mathbf{s}_{A} \oplus \mathbf{z}_{0}$
			\\
			$\bullet$ Bob receives $\mathbf{z}_0$ and computes the key $\mathbf{s}_A$
		\end{algorithm}
	\end{minipage}
	\caption{Alice's and Bob's actions during the application of the fSEBV protocol.}
	\label{fig:Alice's and Bob's Actions in the sSEBV Protocol}
\end{figure}

\section{Examples illustrating the operation of the protocols} \label{sec:Examples Illustrating the Operation of the Protocols}

This section presents and analyzes two small scale but detailed examples in order to illustrate the operation of the fSEBV and sSEBV protocols in practice. The fSEBV and sSEBV protocols were simulated using IBM's Qiskit open source SDK (\cite{Qiskit}). Note that during our tests it was not possible to simulate in Qiskit Alice and Bob being spatially separated or a third party source providing the entangled EPR pairs. So these important assumptions cannot be accurately reflected 
in the simulation and for that reason the examples do not represent a real life environment. As a result Alice and Bob appear in the same circuit. Specifically, Alice's input register consists of the qubits $\ket{q_2 q_1 q_0}$ and her output register is $\ket{q_3}$. Symmetrically, Bob's input register consists of the qubits $\ket{q_6 q_5 q_4}$ and his output register is $\ket{q_7}$. Moreover, the entangled EPR pairs are created by the circuit itself. These facts can be seen in Figures \ref{fig:fSEBV Circuit} and \ref{fig:sSEBV Circuit}, where in the initial stage of the corresponding circuits Hadamard and CNOT gates are used to populate Alice's and Bob's input registers with entangled EPR pairs, exactly as explained in Section \ref{sec:Preliminaries}.

\subsection{Example for the fSEBV protocol}

\begin{figure}[H]
	\centering
	\includegraphics[scale = 0.46, trim = {0 0 0cm 0}, clip]{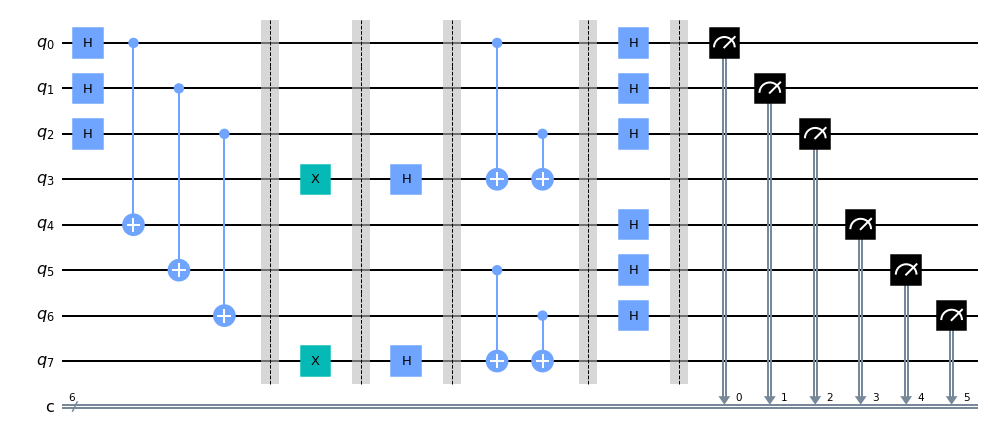}
	\caption{The circuit for the fSEBV protocol.}
	\label{fig:fSEBV Circuit}
\end{figure}

In this example it is assumed that $\mathbf{s}_{A} = 101$ and $\mathbf{s}_{B} = 110$. The resulting circuit in displayed in Figure \ref{fig:fSEBV Circuit}.

\begin{figure}[H]
	\centering
	\includegraphics[scale = 0.2, trim = {0 0 0cm 0}, clip]{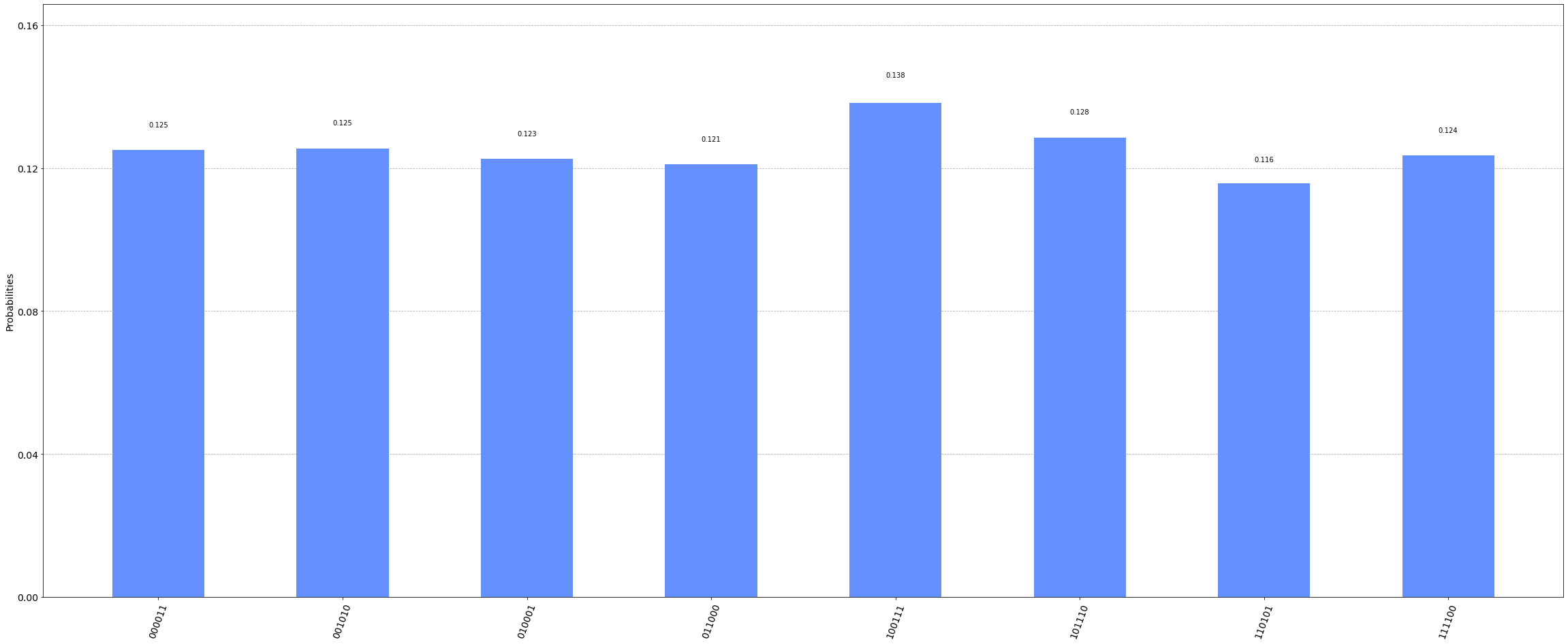}
	\caption{The possible outcomes of the measurement and their corresponding probabilities for the circuit in Figure \ref{fig:fSEBV Circuit}.}
	\label{fig:fSEBV Measurement Outcomes}
\end{figure}

The final measurement by Alice and Bob will produce one of the $8$ outcomes shown in Figure \ref{fig:fSEBV Measurement Outcomes} along with their corresponding probabilities as given by running the simulation for $2048$ shots. A simple inspection of the possible outcomes confirms equation (\ref{eq:QKD BV Final Measurement}). This is because every possible outcome can be written either as $\ket{ \mathbf{z}_{0} }_{A} \ket{ \mathbf{s}_{A} \oplus \mathbf{s}_{B} \oplus \mathbf{z}_{0} }_{B}$ or as $\ket{ \mathbf{s}_{A} \oplus \mathbf{s}_{B} \oplus \mathbf{w}_{0} }_{A} \ket{ \mathbf{w}_{0} }_{B}$, for some, generally different, $\mathbf{z}_{0}, \mathbf{w}_{0} \in \{ 0, 1 \}^n$. Hence, Bob, after measuring (and accepting) the secret key $\mathbf{s}_{A} \oplus \mathbf{s}_{B} \oplus \mathbf{z}_{0}$, just needs to send his secret key $\mathbf{s}_{B} = 110$ to Alice so that she too can derive the secret key.

To avoid any confusion, we clarify that the measurements shown in Figure \ref{fig:fSEBV Measurement Outcomes} give both Alice's and Bob's input registers as $\ket{q_6 q_5 q_4 q_2 q_1 q_0}$.

\subsection{Example for the sSEBV protocol}

\begin{figure}[H]
	\centering
	\includegraphics[scale = 0.46, trim = {0 0 0cm 0}, clip]{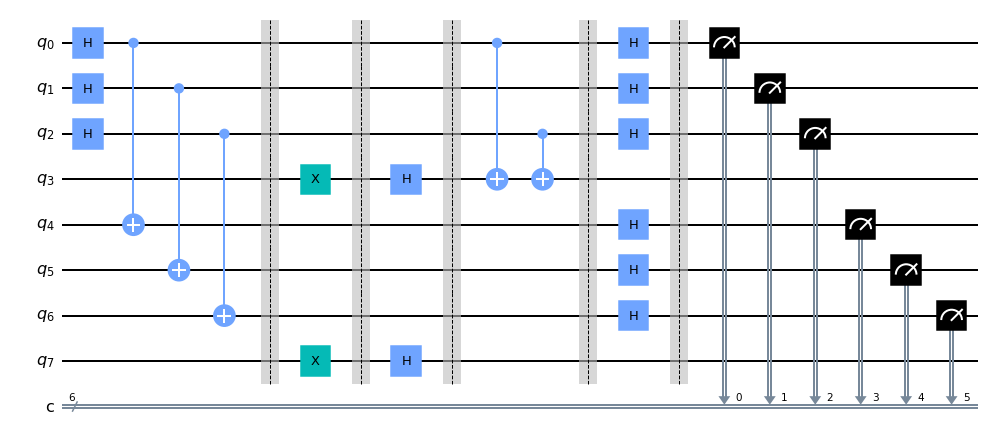}
	\caption{The circuit for the sSEBV protocol.}
	\label{fig:sSEBV Circuit}
\end{figure}

In this example it is assumed that $\mathbf{s}_{A} = 101$ and $\mathbf{s}_{B} = 000$. The resulting circuit in displayed in Figure \ref{fig:sSEBV Circuit}.

\begin{figure}[H]
	\centering
	\includegraphics[scale = 0.2, trim = {0 0 0cm 0}, clip]{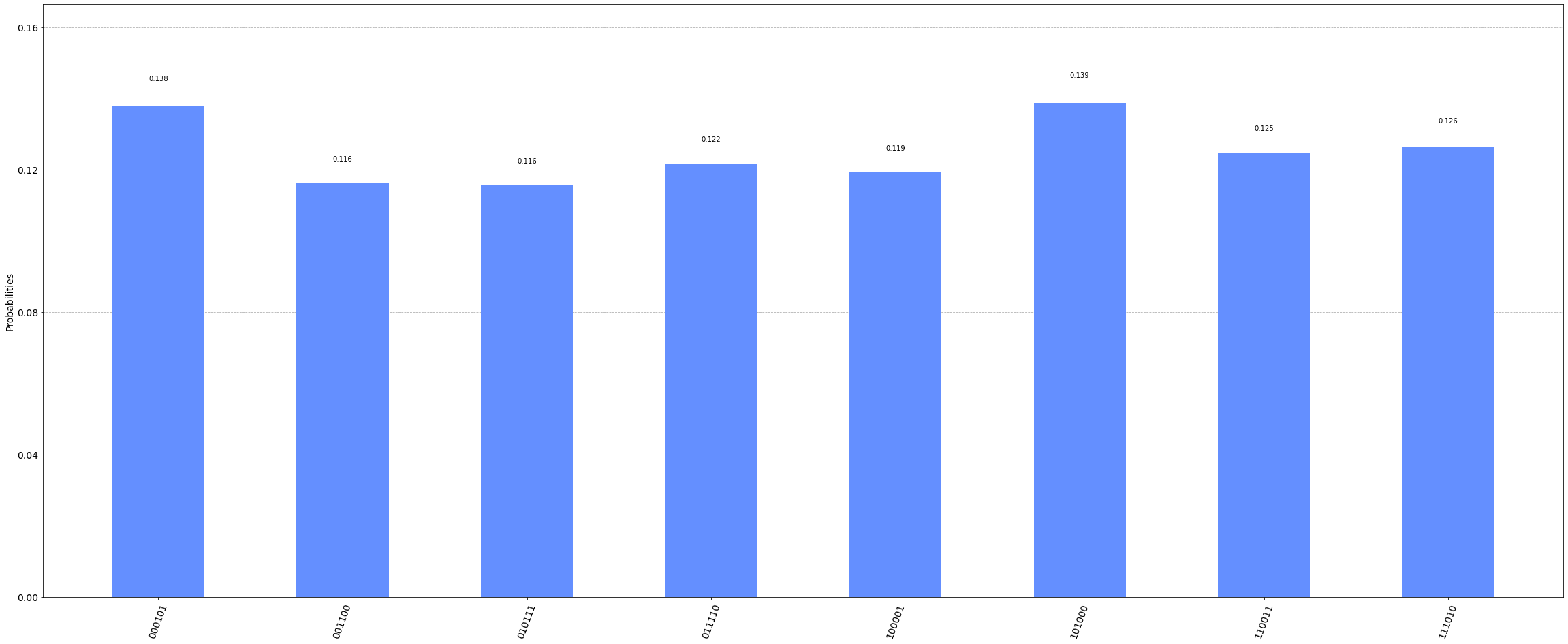}
	\caption{The possible outcomes of the measurement and their corresponding probabilities for the circuit in Figure \ref{fig:sSEBV Circuit}.}
	\label{fig:sSEBV Measurement Outcomes}
\end{figure}

This time the final measurement by Alice and Bob will produce one of the $8$ outcomes shown in Figure \ref{fig:sSEBV Measurement Outcomes} along with their corresponding probabilities as given by running the simulation for $2048$ shots. A noted in the previous case, it suffices to inspect the possible outcomes in order to confirm equation (\ref{eq:QKD BV Final Measurement - II - II}). Now the correct interpretation of the outcomes means viewing them either as $\ket{ \mathbf{z}_{0} }_{A} \ket{ \mathbf{s}_{A} \oplus \mathbf{z}_{0} }_{B}$ or as $\ket{ \mathbf{s}_{A} \oplus \mathbf{w}_{0} }_{A} \ket{ \mathbf{w}_{0} }_{B}$, for some, generally different, $\mathbf{z}_{0}, \mathbf{w}_{0} \in \{ 0, 1 \}^n$. Hence, Alice after making her final measurement and finding a random binary string $\mathbf{z}_{0}$, she just needs to send $\mathbf{z}_{0}$ to Bob. Then Bob will be able to derive Alice's chosen secret key $\mathbf{s}_{A} = 101$.

Again we clarify that the measurement shown in Figure \ref{fig:sSEBV Measurement Outcomes} contain both Alice's and Bob's input registers as $\ket{q_6 q_5 q_4 q_2 q_1 q_0}$.

\section{Conclusions} \label{sec:Conclusions}

Quantum key distribution protocols have surely proved by now that they are a very possible near future reality. Allowing us to harness the power of quantum-mechanics and nature's own laws, in order to achieve a secure key distribution, instead of relying on complex mathematical problems, which base their security on the simple fact, that we have not found an efficient answer to them yet. In this paper, we tried to further expand the field of quantum cryptography, by proposing a novel use for the Bernstein-Vazirani algorithm as a symmetrical entanglement-based QKD protocol, coming in two flavors. In the fully symmetric version the protocol has the ability to create a totally new and original key that both Alice and Bob were initially oblivious of. In the semi-symmetric version, a specific key that is chosen by either Alice or Bob, is securely transmitted to the other party. Additionally, we demonstrated two small scale but comprehensive examples, illustrating the operation of the two protocols in practice, and, finally, we
showcased the protocols strength against an eavesdropping attack by Eve, due to the inherent robustness of entanglement-based protocols against Eve’s attacks as originally described by Ekert and the added security of using extra inputs in order to acquire the final key. We also believe, that the rest of the old quantum algorithms, such as the Deutsch-Jozsa algorithm and Simon’s periodicity algorithm, can all be implemented as a symmetrical entanglement-based QKD protocols, posing a viable suggestion for future work, along with the performance of these proposals against different quantum attacks.

\bibliographystyle{ieeetr}
\bibliography{QKDviaSEBV}

\end{document}